\documentclass[%
 rsi,
 amsmath,amssymb,
preprint,%
]{revtex4-1}

\usepackage{graphicx}% Include figure files
\usepackage{dcolumn}% Align table columns on decimal point
\usepackage{bm}% bold math
\usepackage[utf8]{inputenc}
\usepackage[T1]{fontenc}
\usepackage{mathptmx}
\usepackage{etoolbox}

\makeatletter
\def\@email#1#2{%
 \endgroup
 \patchcmd{\titleblock@produce}
  {\frontmatter@RRAPformat}
  {\frontmatter@RRAPformat{\produce@RRAP{*#1\href{mailto:#2}{#2}}}\frontmatter@RRAPformat}
  {}{}
}%
\makeatother
\begin{document}

%\preprint{AIP/123-QED}

\title[]{Development of a near-5-Kelvin, cryogen-free, pulse-tube refrigerator-based scanning probe microscope}
% Force line breaks with \\
\author{Jun Kasai}
\affiliation{ 
UNISOKU Co., Ltd. Hirakata, Osaka 573-0131, Japan
}%

\author{Tomoki Koyama}
\affiliation{ 
UNISOKU Co., Ltd. Hirakata, Osaka 573-0131, Japan
}%

\author{Munenori Yokota}
\affiliation{ 
UNISOKU Co., Ltd. Hirakata, Osaka 573-0131, Japan
}%

\author{Katsuya Iwaya}
%\email[Corresponding author:]{katsuya\_iwaya@unisoku.co.jp}
\affiliation{ 
UNISOKU Co., Ltd. Hirakata, Osaka 573-0131, Japan
}%

%\date{\today}

\begin{abstract}
We report the design and performance of a cryogen-free, pulse-tube refrigerator (PTR)-based scanning probe microscopy (SPM) system capable of operating at the base temperature of near 5~K. We achieve this by combining a home-made interface design between the PTR cold head and the SPM head, with an automatic gas-handling system. The interface design isolates the PTR vibrations by a combination of polytetrafluoroethylene and stainless-steel bellows, and by placing the SPM head on a passive vibration isolation table via two cold stages that are connected to thermal radiation shields using copper heat links. The gas-handling system regulates the helium heat-exchange gas pressures, facilitating both the cool down to and the maintenance of the base temperature. We discuss the effects of each component using measured vibration, current-noise, temperature, and pressure data. We demonstrate that our SPM system performance is comparable to known liquid-helium-based systems with the measurements of the superconducting gap spectrum of Pb, atomic-resolution scanning tunneling microscopy image and quasiparticle interference pattern of Au(111) surface, and non-contact atomic force microscopy image of NaCl(100) surface. Without the need for cryogen refills, the current SPM system enables uninterrupted low-temperature measurements.
\end{abstract}

\maketitle

\section{INTRODUCTION}

Scanning tunneling microscopy (STM) and atomic force microscopy (AFM) are powerful analytical techniques that probe surface phenomena down to the atomic scale for a wide range of materials and experimental conditions\cite{Wiesendanger}.  As such, STM and AFM experiments are often performed at cryogenic temperatures to shut down all thermally activated processes—a condition that minimizes thermal drift, providing stability to the probe-sample junction; molecular and atomic migrations, providing stability both to the probe apex and to the sample surface; and thermal broadening, enhancing both spatial and energy resolutions. In STM, the low-temperature conditions are critical for tunneling spectroscopy measurements, making it possible to explore the physical and chemical phenomena with high energy resolution; in non-contact AFM, the thermal frequency noise that fluctuates the frequency shifts are substantially reduced. Owing to these advantages, many atomic-scale phenomena have been investigated using various scanning probe techniques at low temperatures\cite{Wiesendanger, Morita}. 

In the past decades, the severe shortage of helium\cite{Nuttall} has made it increasingly difficult to conduct low-temperature scanning probe microscopy (SPM) experiments. While the use of low helium boil-off cryostats\cite{Okamoto, Trofimov, Unisoku} can alleviate the problem, the ultimate solution lies in the integration of a cryogen-free refrigerator into the SPM system itself. Such integration would not only eliminate the dependency on scarce helium, but it would also enable experiments to be carried out uninterrupted without the need for cryogen refills. To this end, several SPM systems that have integrated cryogen-free refrigerators using elaborate mechanical vibration isolations have been reported\cite{denHaan, Hackley, Zhang, Pabbi, Meng, Chaudhary} and made commercially available\cite{Omicron, RHK}. However, realizing an instrument with both a base temperature close to 4.2 ~K, and high-resolution measurement capabilities similar to liquid-helium-based SPM systems has remained elusive. 

In this paper, we present a cryogen-free ultrahigh vacuum (UHV) SPM system capable of operating at the base temperature of near 5 K using an integrated pulse-tube refrigerator (PTR). We achieve this by combining a home-made interface between the PTR cold head and the SPM head with a new automatic gas-handling system. The home-made interface is designed to suppress resonant coupling of the mechanical vibrations from the PTR to the SPM. The gas-handling system automatically regulates the pressure of helium heat-exchange gas to cool the cryostat down to and keep the cryostat at the base temperature. We discuss the effects of vibration isolation using systematic vibration noise measurements. We demonstrate the performance of the cryogen-free SPM by measuring the superconducting gap of a Pb surface, by performing STM experiments on a Au(111) surface, and by performing non-contact AFM experiments on a NaCl(100) surface. We show that both the base temperature and the data quality from our cryogen-free SPM system are comparable to conventional liquid-helium-based SPM systems.

\section{SYSTEM OVERVIEW}

%Fig.~\ref{SystemOverview}%
\begin{figure*}
\includegraphics[width=16cm]{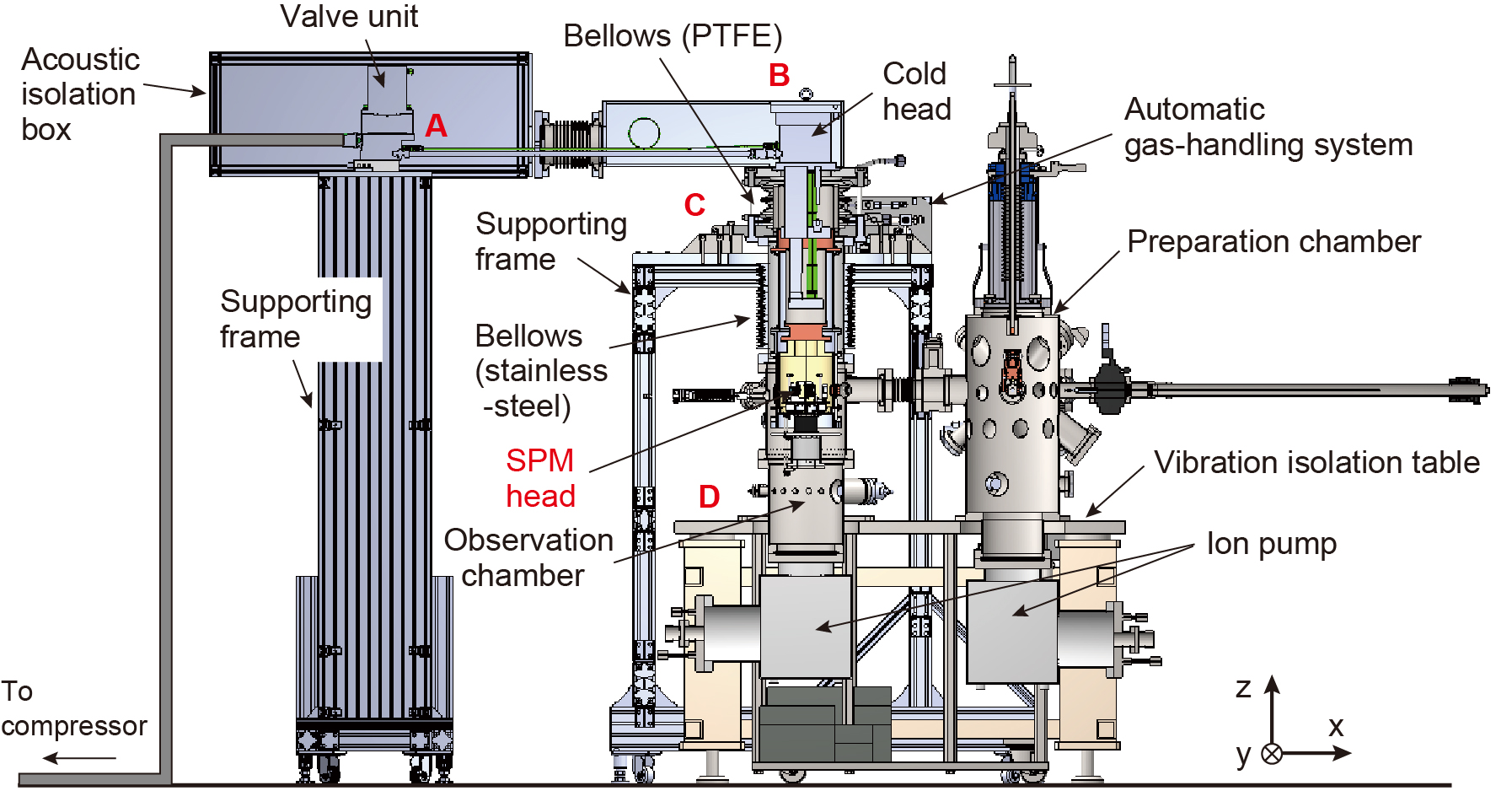}
\caption{\label{SystemOverview}Schematic of the entire cryogen-free low-temperature SPM system. The letters A–D denote the locations where vibration noise spectra are measured.}
\end{figure*}

Figure~\ref{SystemOverview} shows the schematic of the cryogen-free low-temperature SPM, located at the first floor of the factory in UNISOKU Co., Ltd. amongst several other SPM systems. The SPM system in the figure is not isolated by sound-proof walls, and is composed of three UHV chambers: a load-lock chamber for tip and sample exchange, a tip and sample preparation chamber, and a SPM observation chamber. The three chambers are mounted on a passive vibration isolation table (Stable 66-1209-8829-CE, Kurashiki Kako, Japan). The preparation and observation chambers are pumped by two ion pumps (240 L/s and 125 L/s), each equipped with a titanium sublimation pump. The base pressure of the observation chamber is $2\times10^{-8}$~Pa at room temperature and $6\times10^{-9}$~Pa when the SPM head is cooled down.

Next, we discuss the coupling of the PTR to the SPM observation chamber. The PTR (RP-082B2S, Sumitomo Heavy Industries, Japan) consists of a water-cooled compressor (F-70LP, Sumitomo Heavy Industries), a valve unit, and a cold head attached to two cooling stages (referred to as 1st and 2nd cooling stages in Fig. 2). The cooling power of the PTR is 35~W at 45~K in the 1st cooling stage and 0.9~W at 4.2~K in the 2nd cooling stage. The power line frequency of 60 Hz is used in this study. The compressor is encased in a home-made acoustic isolation box and is located $\sim 7$~m away from the SPM system for sound isolation. The length of the flexible tubes interconnecting the compressor and the valve unit is 20~m. Ideally, the compressor should be located in a separate room for optimum sound isolation. The valve unit is mounted on a purposefully rigid and heavy supporting frame, and connected to the PTR cold head via two stainless-steel tubes and one flexible tube. The two stainless-steel tubes are fixed together to reduce vibration transmission from the valve unit to the cold head. The valve unit, the cold head, and the tubes are encased in an acoustic isolation box.

%Fig.~\ref{Interface}%
\begin{figure}
\includegraphics[width=8cm]{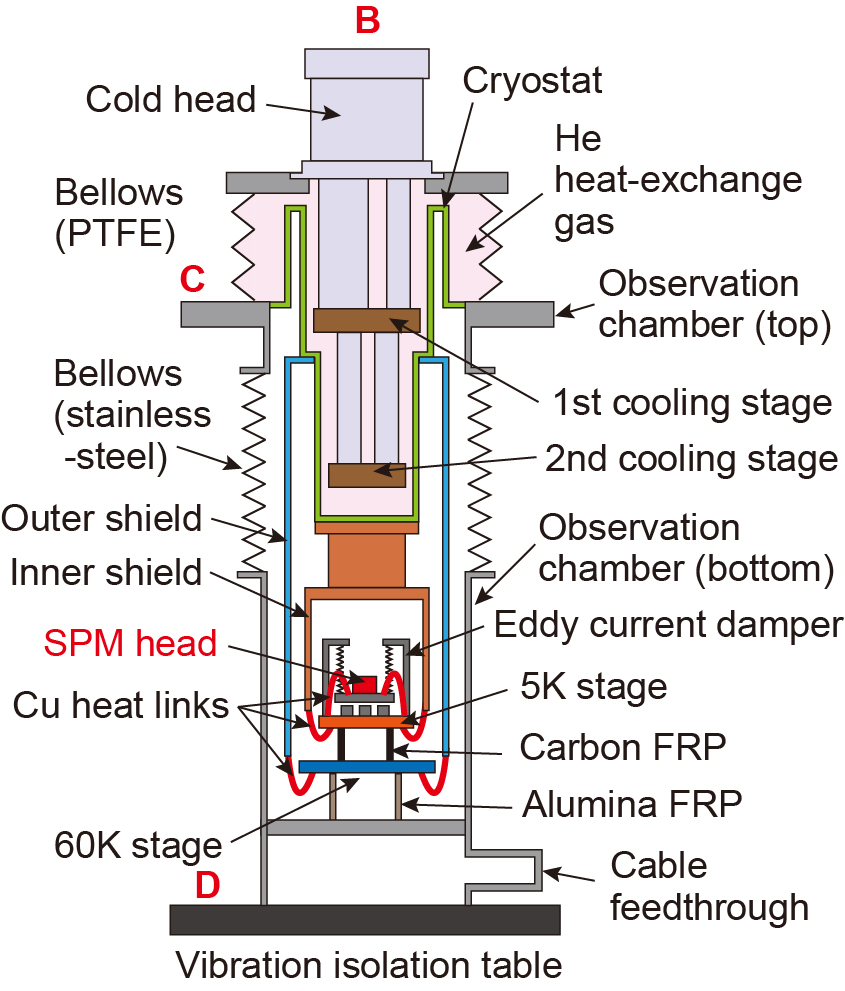}
\caption{\label{Interface} Schematic of the interface between the PTR and the SPM head. The area enclosed by the cold head, the PTFE bellows, and the cryostat is filled with helium that serves as heat-exchange gas between the cooling stages and the bottom of the cryostat. The observation chamber is kept in UHV condition. The letters B–D denote the locations where vibration noise spectra are measured.}
\end{figure}

The cold head is attached at the top of the observation chamber using polytetrafluoroethylene (PTFE) bellows to suppress vibration transmissions to the chamber (Fig.~\ref{Interface}). The top of the observation chamber is mounted on its own supporting frame separated from the vibration isolation table (Fig.~\ref{SystemOverview}). To further suppress the vibration transmission from the cold head, stainless-steel bellows are placed between the top and bottom part of the observation chamber (Fig.~\ref{Interface}).

The space enclosed by the cold head, the PTFE bellows, and the cryostat (area indicated by the arrow labeled "He heat-exchange gas" in Fig.~\ref{Interface}) houses the 1st and 2nd cooling stages of the PTR, and is filled with helium gas for heat exchange between the cooling stages and the cryostat. The pressure of helium heat-exchange gas during cooling is automatically controlled by the gas-handling system which will be described in detail later. We note that this design is suitable for the baking of the observation chamber because components that are not heat-resistant (i.e., the cold head, the two cooling stages, and the PTFE bellows and O-rings) can be easily removed. 

Next, we discuss the setup of the SPM head. The SPM head is surrounded by two layers of thermal radiation shields (Fig.~\ref{Interface}). The outer aluminum (purity $\geq 99.7$~\%) shield is attached to the cryostat near the 1st cooling stage, and the inner oxygen-free copper shield is directly attached to the bottom of the cryostat. The outside surface of the inner shield is gold-plated. The outer radiation shield has four windows and the inner radiation shield has three through-holes and a window aligned with the UHV chamber flanges near the SPM head. Two pairs of outer window and inner through-hole are aligned for the observation of the tip apex and sample surface. A pair of outer window and inner window is aligned for
transferring the tip and sample to and from the SPM head. Another pair of outer window and inner through-hole is aligned for locking the SPM head during the tip/sample transfer. Each window can be closed by its own shutter.

%Fig.~\ref{SPMhead}%
\begin{figure}
\includegraphics[width=8.5cm]{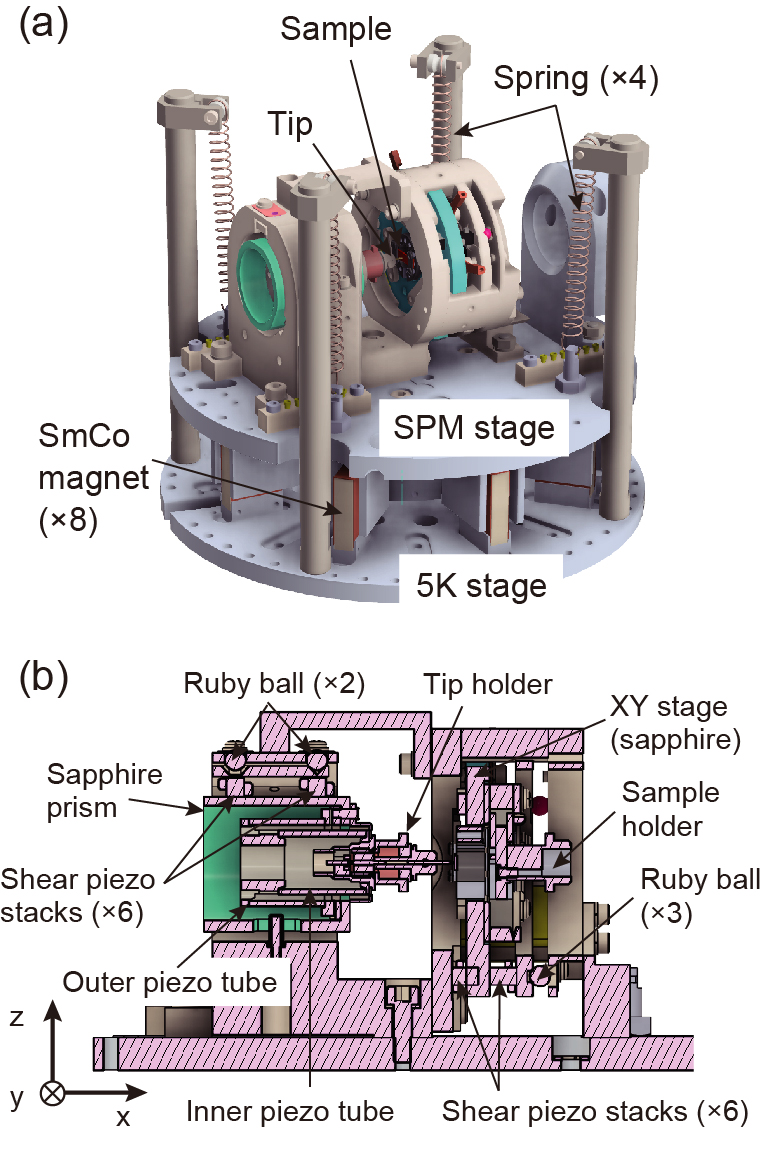}
\caption{\label{SPMhead}(a) Three-dimensional illustration of the SPM unit. An eddy current damper is used between the SPM stage and the 5~K stage for vibration isolation. The copper heat links between SPM and 5~K stages and electrical wirings are not shown for clarity. (b) Cross-section of the SPM head. A tip holder for STM is shown in both figures. The tip and sample are aligned horizontally in the x direction.}
\end{figure}

The SPM head is mounted on two cold stages. Figure 2 shows both the 5~K and the 60~K stages that are connected to the inner and to the outer radiation shield via copper heat links, respectively. The copper heat links not only cool down the cold stages, but they also dampen vibration noise from the cold head. The copper heat links connecting the outer shield and the 60~K stage are composed of twenty-four copper cables each braded from 3,200 copper wires (diameter $\phi = 0.05 $~mm, and length $L = 55$~mm for each wire). The total cross-sectional area of  the heat links is $\sim 150~\rm{mm}^2$. The copper heat links connecting the inner shield and the 5~K stage are composed of twenty-four copper cables each braded from 1,600 copper wires ($\phi = 0.05$~mm, $L = 45$~mm for each wire). The total cross-sectional area of the heat links is $\sim 75~\rm{mm}^2$. 

The 5 K and 60 K stages are supported by a carbon and an alumina fiber-reinforced plastic (FRP) cylinder, respectively (Fig.~\ref{Interface}). We choose these materials for their low thermal conductivities and large Young’s moduli, which promote the efficient cooling and enhance both the thermal and the vibration isolations of the SPM head. The alumina FRP has been used as supporting rods in the ultra-low-vibration cryocooler system for the gravitational wave detector\cite{Ikushima}. The carbon FRP is known to exhibit sufficiently low thermal conductivity at low temperatures\cite{Takeno} with larger Young’s modulus than the alumina FRP. 

The SPM head is further vibrationally isolated by an eddy current damper using four stainless-steel springs and eight samarium-cobalt (SmCo) magnets as shown in Fig.~\ref{SPMhead}(a). Sixteen stranded copper cables consisting of $\sim$~380 copper wires with a diameter $\phi = 0.05$~mm and the length $L = 40$~mm are used as heat links between the 5~K stage and the SPM stage (total cross-sectional area is $\sim 12~\rm{mm}^2$). 

The SPM head is built according to the UNISOKU USM1300 system, with some component modifications to increase rigidity. The sample XY stage and the prism in the Z stage for tip coarse approach, both moved by stick and slip motion of shear piezo stacks, are made of sapphire, also for rigidity [Fig.~\ref{SPMhead}(b)]. To enhance the SPM head resonance frequencies even when a heavy tip holder ($\sim 1.8$~g) is mounted on top of the piezo tube for scanning, we use the double piezo tube design, where both the inner and outer tubes are used for XY scanning, and the outer tube is used for the feedback Z motion [Fig.~\ref{SPMhead}(b)]. The configuration of the electrodes in the piezo tubes is shown in Appendix A. The maximum scan range is $1.5~\rm{\mu m} \times 1.5~\rm{\mu m}$ at low temperatures. The lowest resonance frequencies at the Z stage and at the XY stage are measured to be $\sim 4$~kHz and $\sim 3.8$~kHz, respectively. The tip and sample are aligned horizontally in the x direction, as defined in Fig.~\ref{SystemOverview}.

We use home-made vacuum-insulated cables as reported by Mykkänen $\it{et~al}$.\cite{Mykkanen} ($L = 17$~cm) to carry the tunneling-current and the AFM signals at the 5~K stage. The vacuum-insulated cable consists of a stainless tube ($\phi = 2.8$~mm, thickness of 0.18~mm) and a CuNi wire ($\phi = 0.1$~mm) and is terminated with SMPM connectors. The total length of the coaxial cable between the tip and the input of the preamplifier located outside the observation chamber is $\sim 50$~cm.

\section{HELIUM GAS-HANDLING SYSTEM}

%Fig.~\ref{GasHandlingSystem}%
\begin{figure*}
\includegraphics[width=16cm]{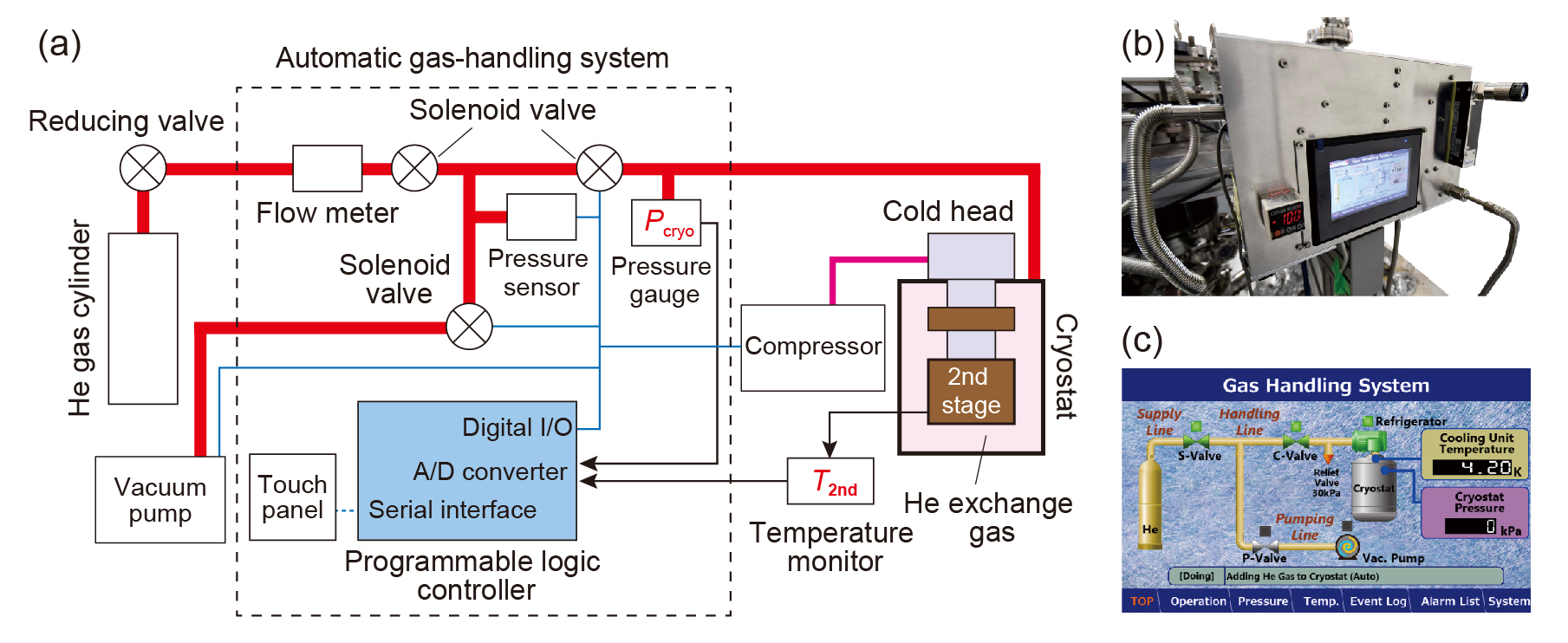}
\caption{\label{GasHandlingSystem} (a) Schematic diagram of the automatic gas-handling system. (b) Photograph of the control panel. (c) Screenshot of the touch panel.}
\end{figure*}

The regulation of the helium heat-exchange gas pressure in the cryostat is important to both the cooling down to and the keeping of a stable base temperature. To this end, we have developed an automatic bimodal helium gas-handling system based on the programmable logic controller [Figs.~\ref{GasHandlingSystem}(a) and (b)]. First, to cool down the cryostat from room temperature to the base temperature of $\sim 5$~K, we need to add the adequate amount of helium heat-exchange gas in the cryostat. The program to cool down from room temperature to the base temperature is referred to as “cool-down mode” hereafter. Second, after cooling down, the helium heat-exchange gas needs to be occasionally refilled to keep the base temperature because the SPM head warms up during tip/sample transfers, expanding and releasing helium gas through the cryostat relief valve. This second program mode is referred to as “maintenance mode” hereafter. The gas-handling system developed in this study enables us to cool the cryostat down to the base temperature and keep the base temperature stable just by pressing a start button on the touch-screen control panel [Fig.~\ref{GasHandlingSystem}(c)].

%Fig.~\ref{CoolingMode}%
\begin{figure}
\includegraphics[width=8.5cm]{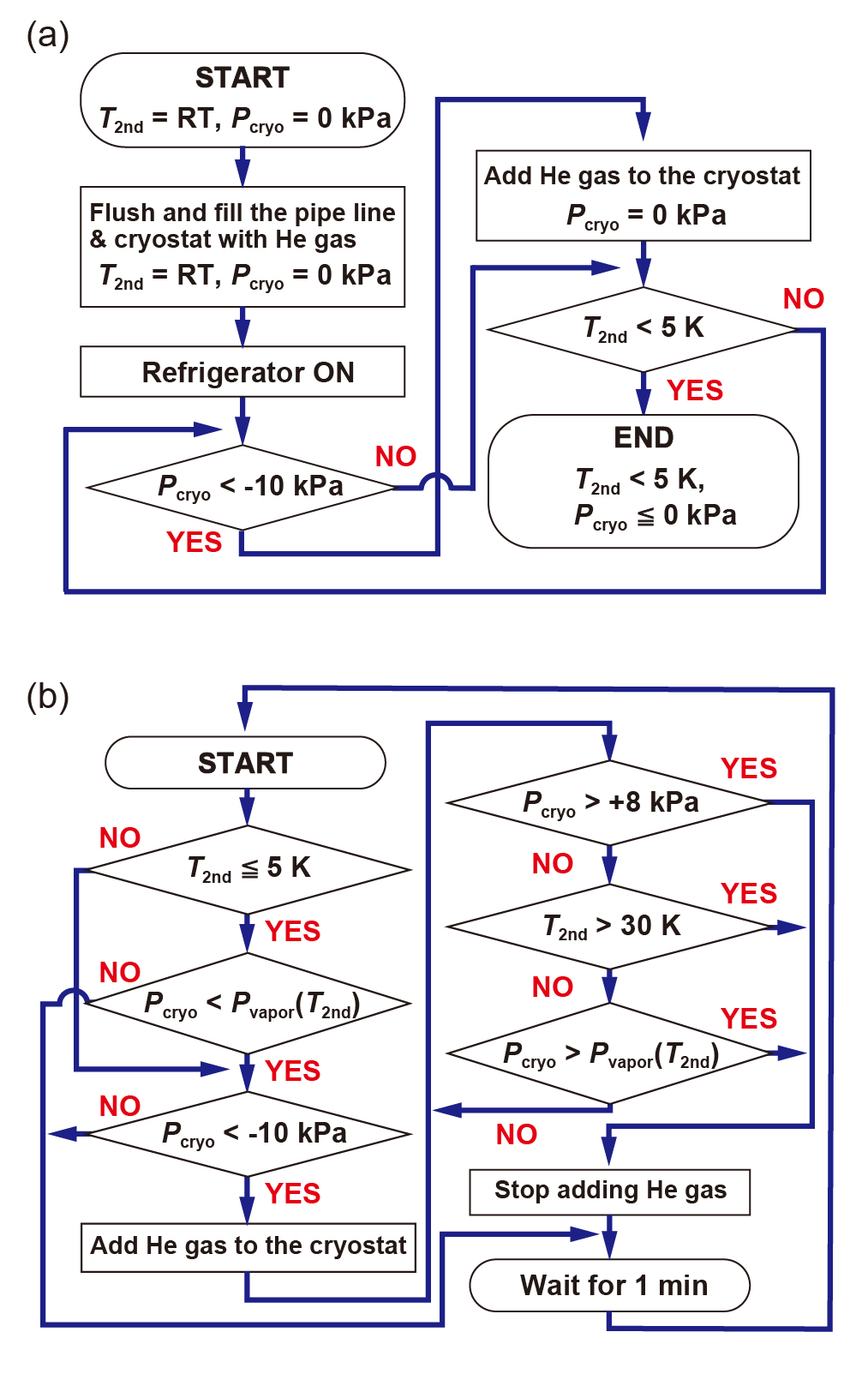}
\caption{\label{CoolingMode}(a) Flowchart describing the cool-down mode from room temperature to 5~K. (b) Flowchart describing the maintenance mode to keep the base temperature of $\sim 5$~K at the 2nd cooling stage of the PTR.}
\end{figure}

Figure~\ref{CoolingMode}(a) describes each process executed during the cool-down mode. The program starts by flushing and filling all plumbing lines and the cryostat with helium gas. After the pressure in the cryostat ($P_{\rm{cryo}}$) reaches atmospheric pressure, the PTR is turned on. In this study, we describe pressure in relative terms, defining the latter condition of 1~atm of helium as $P_{\rm{cryo}}=0~\rm{kPa}$. 

As $P_{\rm{cryo}}$ drops with temperature, helium gas must be added to the cryostat to prevent the cooling rate from slowing down. The threshold pressure that triggers the helium-gas addition is set to the empirically chosen 10~kPa lower than atmospheric pressure ($P_{\rm{cryo}} < -10~\rm{kPa}$). When the condition $P_{\rm{cryo}} < -10~\rm{kPa}$ is met, the program opens the solenoid valves in Fig.~\ref{GasHandlingSystem}(a) and fills the cryostat with helium gas up to $P_{\rm{cryo}}=0~\rm{kPa}$. This process repeats until the temperature at the 2nd cooling stage ($T_{\rm{2nd}}$) drops down below 5 K. The end of the cool-down mode is reached
when both conditions $T_{\rm{2nd}} < 5~\rm{K}$ and $P_{\rm{cryo}} \leq 0~\rm{kPa}$ are met.

After completion of the cool-down mode, the program switches to the maintenance mode to keep $T_{\rm{2nd}} \leq 5~\rm{K}$. Figure~\ref{CoolingMode}(b) describes how the program evaluates when to start/stop the addition of helium gas into the cryostat. At $T_{\rm{2nd}} \leq 5~\rm{K}$, helium gas is added when $P_{\rm{cryo}}$ is both lower than the vapor pressure of helium at $T_{\rm{2nd}}$ [$P_{\rm{cryo}} < P_{\rm{vapor}}(T_{\rm{2nd}})$], and lower than the threshold pressure of $-10$~kPa ($P_{\rm{cryo}} < -10~\rm{kPa}$). At $T_{\rm{2nd}} > 5~\rm{K}$, only the latter condition ($P_{\rm{cryo}} < -10~\rm{kPa}$) is used to decide whether helium gas needs to be added or not because the vapor pressure of helium is calculated only at $T_{\rm{2nd}} \leq 5~\rm{K}$. 

Upon starting the helium-gas addition, the program evaluates when to stop the addition. For this purpose, we set three following conditions: $P_{\rm{cryo}} > +8~\rm{kPa}$, $T_{\rm{2nd}} > 30~\rm{K}$, and $P_{\rm{cryo}} > P_{\rm{vapor}}(T_{\rm{2nd}})$ [Fig.~\ref{CoolingMode}(b)]. When any one of these three conditions is met, the solenoid valves immediately close to stop the helium-gas addition. The first two conditions are designed to be met by either the $P_{\rm{cryo}}$ and $T_{\rm{2nd}}$ surges during the sample/tip transfers. The threshold values of 8~kPa and 30~K are empirically chosen. The third condition $P_{\rm{cryo}} > P_{\rm{vapor}}(T_{\rm{2nd}})$ is designed to be met during a long-term SPM measurement when $T_{\rm{2nd}} \leq 5~\rm{K}$ occurs. Upon closing the solenoid valves to stop adding helium gas, the program waits for one minute before repeating the entire procedure.

\section{COOLING PERFORMANCE}

%Fig.~\ref{CoolingTest}%
\begin{figure}
\includegraphics[width=8.5cm]{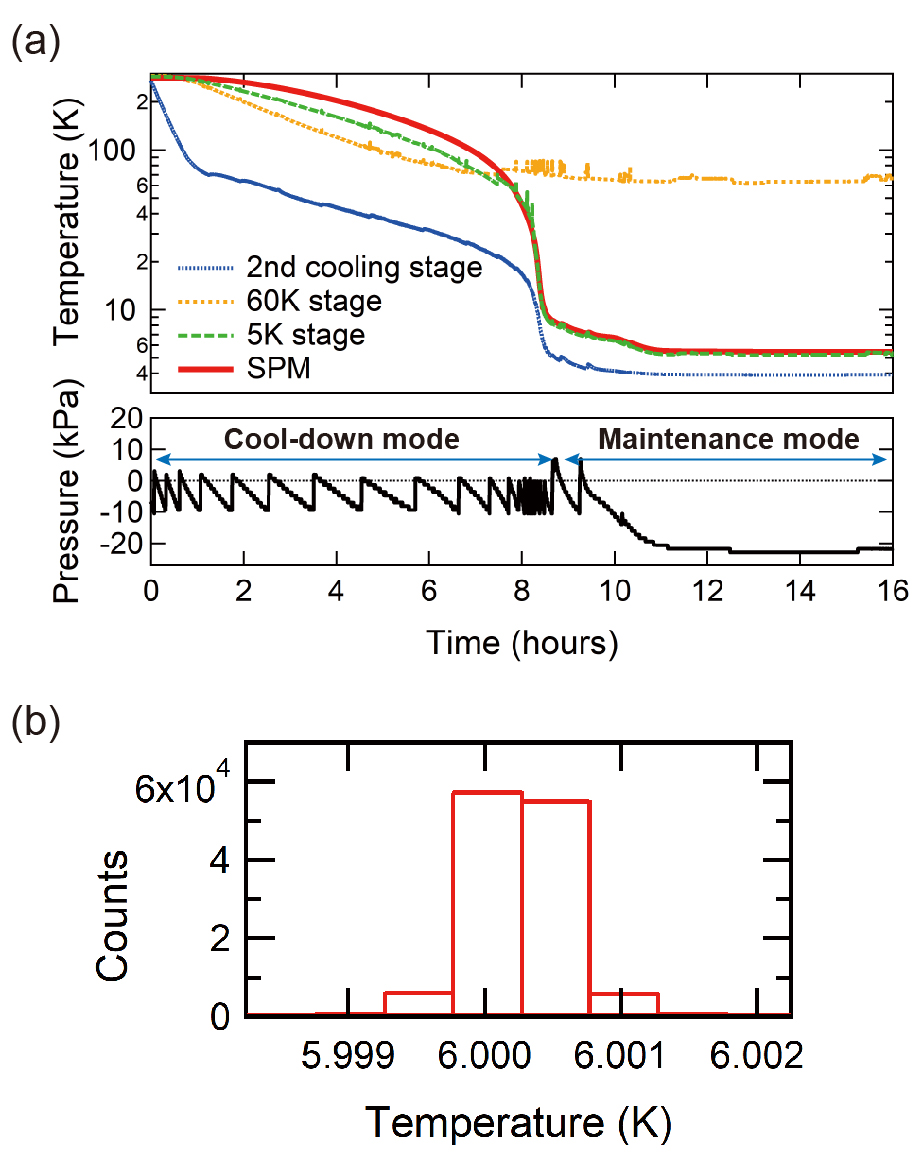}
\caption{\label{CoolingTest}(a) (Top) Typical cool-down temperature curves measured at the 2nd cooling stage, 60~K stage, 5~K stage and SPM head shown in Fig.~\ref{Interface}. (Bottom) Corresponding cryostat pressure controlled by the gas-handling system. The atmospheric pressure corresponds to 0~kPa. (b) Histogram of the SPM head temperature controlled to stabilize at 6~K for three days.}
\end{figure}

Figure~\ref{CoolingTest}(a) shows the typical temperature curves at the 2nd cooling stage, 60~K stage, 5~K stage and SPM head (top panel), as well as the corresponding $P_{\rm{cryo}}$ curve (bottom panel) during the cool-down and the maintenance modes. It takes $\sim 9$~hours to cool the 2nd cooling stage from room temperature down to $\sim 5$~K. While the temperatures at the 60~K and the 5~K stages are sensitive to the variations in $P_{\rm{cryo}}$, the temperature at the SPM head is not. After twelve hours, the base temperatures are reached at the 2nd cooling stage, 60~K stage, 5~K stage and SPM head at 3.9~K, 63~K, 5.2~K and 5.5~K, respectively. The small temperature difference between the 5~K stage and the SPM head (0.3~K) indicates good thermal contact by the heat links. After reaching the base temperature, the helium-gas addition to the cryostat becomes a rare event that only occurs during tip/sample transfers.

The base temperature fluctuates over a 24-hr period depending on room temperature, typically in the range of 100~mK at the SPM head. We minimize this temperature fluctuation using a feedback-controlled heater attached to the SPM stage. As shown in Fig.~\ref{CoolingTest}(b), the temperature fluctuation at 6~K for three days is reduced to within $\pm1~\rm{mK}$, which is stable for long-term SPM measurements.

\section{VIBRATION MEASUREMENTS}

The interface structure connecting the PTR and the SPM head in Fig.~\ref{Interface} indicates two routes that can transmit vibrations from the cold head to the SPM head. The first route transmits through the PTFE bellows, the cryostat, the inner and outer shields, and the heat links between the radiation shields and the cold stages. For this route, both the PTFE bellows and the heat links are the key components that suppress vibration transmission. The second route transmits through the PTFE bellows, the stainless-steel bellows, the vibration isolation table, and the two FRP cylinders. For this route, since the SPM head is placed on the vibration isolation table via the 5 K and 60 K stages, suppressing vibration transmission using the vibration isolation table is critical. 

%Fig.~\ref{VibrationNoise}%
\begin{figure}
\includegraphics[width=8.5cm]{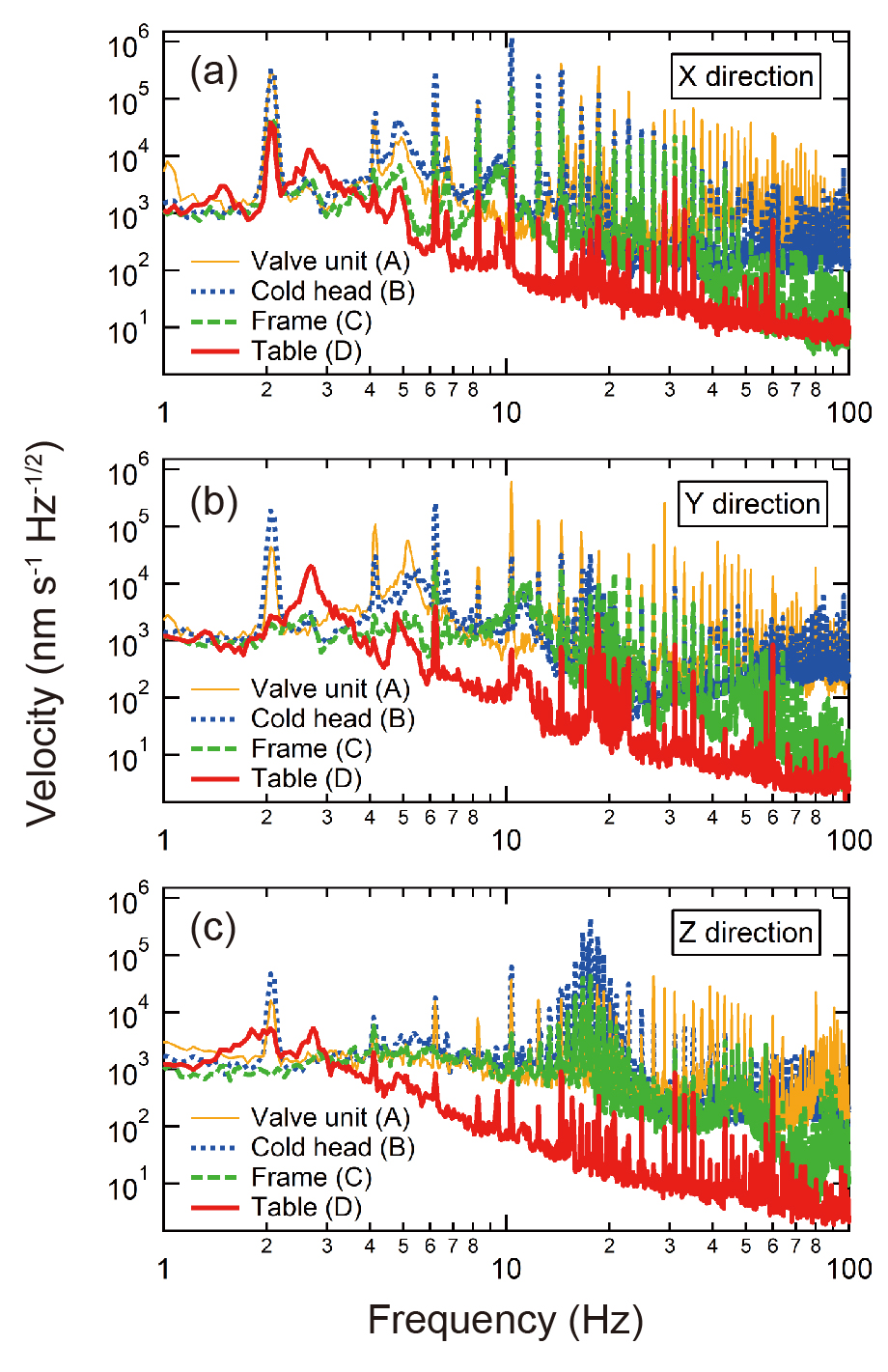}
\caption{\label{VibrationNoise} (a-c) Vibration spectra at valve unit (A), cold head (B), frame (C) and vibration isolation table (D) in the x, y and z directions, respectively. Refer to Fig.~\ref{SystemOverview} for the depictions of letters A–D. All measurements are acquired both the PTR and the passive vibration isolation table in operation. The scale in y axis is the same in all figures. }
\end{figure}

To analyze the vibration transmissions, we measure the vibration velocity spectra along the x, y and z direction on the valve unit (A), the cold head (B), the supporting frame (C) and the vibration isolation table (D) during PTR operation (see Fig.~\ref{SystemOverview} for locations of A–D). The vibration spectra are measured using a commercial accelerometer (NP-7310, Onosokki, Japan), its amplifier (PS-1300, Onosokki, Japan), and a Nanonis SPM controller (SPECS GmbH, Germany). 
%Fig.~\ref{VibrationNoise2}%
\begin{figure}
\includegraphics[width=8.5cm]{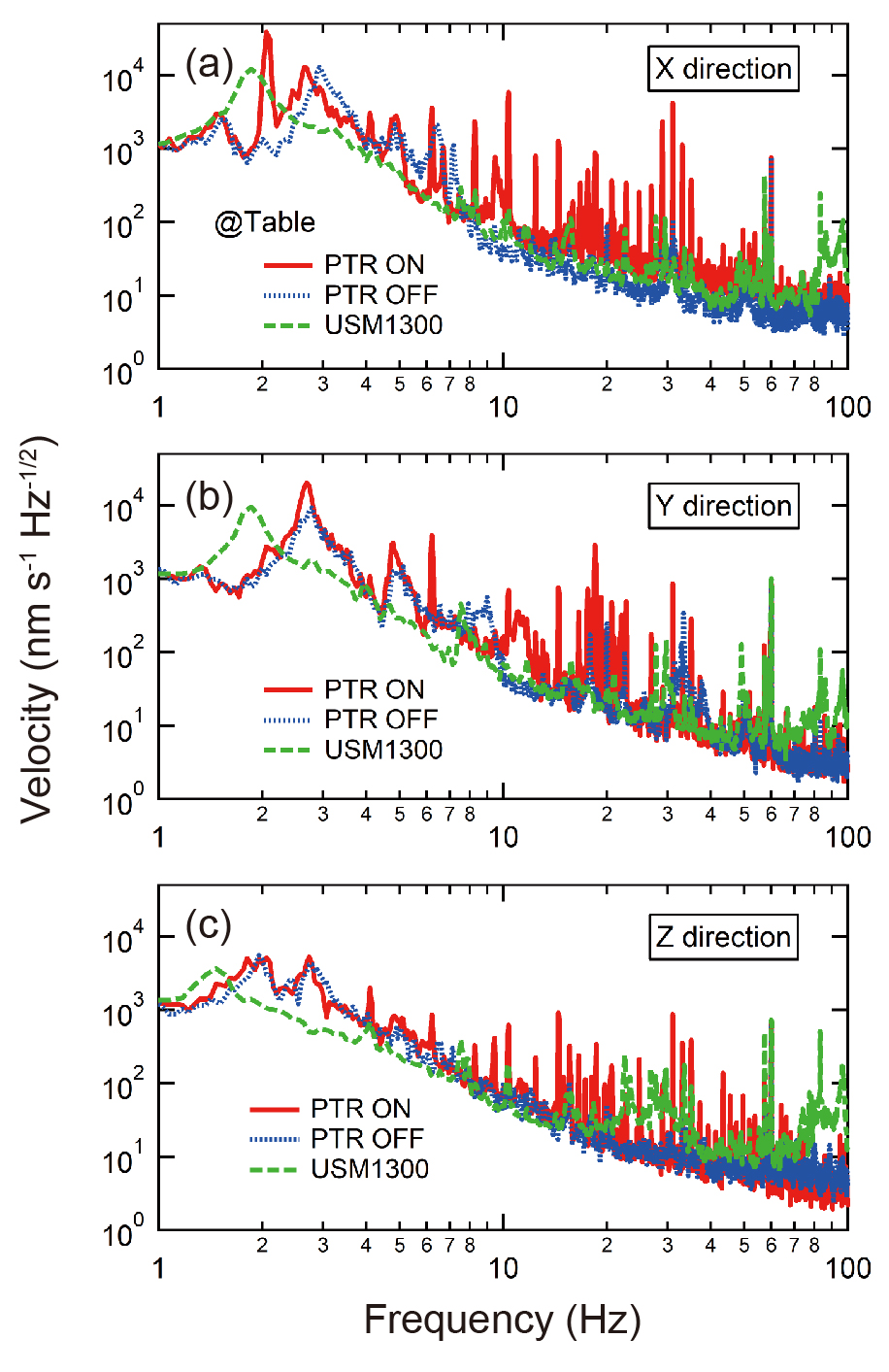}
\caption{\label{VibrationNoise2} (a-c) Effects of the PTR on velocity spectra at the passive vibration isolation table (D in Fig.~\ref{SystemOverview}) in the x, y and z directions, respectively. For comparison, the velocity spectra at the passive vibration isolation table in a conventional SPM system with a liquid helium cryostat (UNISOKU USM1300), placed next to the PTR-based SPM, are plotted. The scale in y axis is the same in all figures.}
\end{figure}

We first compare the vibration spectra acquired at the valve unit and the cold head (light-solid and dotted lines in Fig.~\ref{VibrationNoise}, respectively). The noise spectra at the valve unit exhibit similar series of peaks in all three directions. The peaks at 2~Hz observed in all three directions correspond to the fundamental frequency of the PTR. The series of peaks at higher frequencies are its higher harmonics. Since the cold head and the valve unit are mechanically connected along the x direction (Fig.~\ref{SystemOverview}), the peak at 2~Hz in the x direction is most striking. In addition, several broad peak structures at 5, 20, and between 30 and 50~Hz are observed in all three directions. 

At the cold head, a series of peaks between 30 and 90~Hz and the broad peaks between 30 and 50~Hz are effectively reduced in all directions. This is because the two stainless tubes connecting the valve unit and the cold head are rigidly fixed together. In contrast, several peaks are enhanced or newly appear at the cold head, including the peak at 2~Hz and the broad peak near 20~Hz in the y and z directions, and a broad peak structure near 10~Hz in the x and y directions. We also find that compared to the x and y directions, the vibration spectrum in the z direction at the cold head exhibits a qualitatively different feature, as represented by the broad peak near 20~Hz [Fig.~\ref{VibrationNoise}(c)]. We attribute these peaks to characteristic vibration modes of the cold head. 

Next, we discuss the effects of the PTFE bellows by comparing the vibration spectra acquired at the supporting frame with those at the cold head (dashed and dotted lines in Fig.~\ref{VibrationNoise}, respectively). The PTFE bellows are clearly effective at suppressing the noise peak at 2~Hz in all directions; in particular, the vibration noise spectra at the frame show no peak at 2~Hz in the y and z directions [Fig.~\ref{VibrationNoise}(b, c)]. Similarly, other peaks at 5 Hz in the x and y directions, above 40~Hz in all directions, and the broad peak at 20~Hz in the z direction are all substantially reduced at the supporting frame.

Next, we compare the vibration spectra acquired at the vibration isolation table with those at the frame (dark-solid and dashed lines in Fig.~\ref{VibrationNoise}, respectively). While many peaks remain visible, the vibration noise levels above 4~Hz are substantially suppressed in all directions at the vibration isolation table. This suppression is mainly due to the stainless-steel bellows and the heavy vibration isolation table (Fig.~\ref{Interface}). The heat links and the FRP cylinders might also contribute to the noise suppression. Below 4~Hz, however, the noise levels are not improved; the noise peak at 2~Hz at the vibration isolation table in the x direction remains unchanged compared to that at the frame. A broad peak structure near 3~Hz appears in all directions. We attribute the enhancement of vibration near 3~Hz to the resonance vibration of the passive vibration isolation damper rather than the vibration transmission from the PTR. 

We examine this effect by comparing noise spectra at the vibration isolation table with the PTR turned on and off (solid and dotted lines in Fig.~\ref{VibrationNoise2}, respectively), and find that the noise peaks at 2-3~Hz in all directions (except the sharp peak at 2~Hz in the x direction) remain almost unchanged before and after the PTR is turned on. This observation indicates that those noise peaks originate from the passive vibration isolation damper rather than from the PTR. A series of peaks between 10 and 70~Hz are possibly caused by noise from the PTR suggesting insufficient isolation by the acoustic isolation box.

We also compare the vibration noise spectra at the vibration isolation table with those of a liquid-helium-based SPM system (USM1300, UNISOKU) placed next to the cryogen-free SPM system (solid and dashed lines in Fig. 8, respectively). The noise floor of the cryogen-free SPM system is comparable to that of the liquid-helium-based SPM system and even lower above 50~Hz in all directions. We conclude that while many peaks remain observable (for instance, at 2~Hz in the x direction and a series of peaks between 10 and 70~ Hz in all directions), the vibrational-noise levels at the vibration isolation table in the cryogen-free SPM system are effectively suppressed for SPM measurements.

\section{TUNNELING-CURRENT NOISE MEASUREMENTS}

%Fig.~\ref{CurrentNoise}%
\begin{figure}
\includegraphics[width=8.5cm]{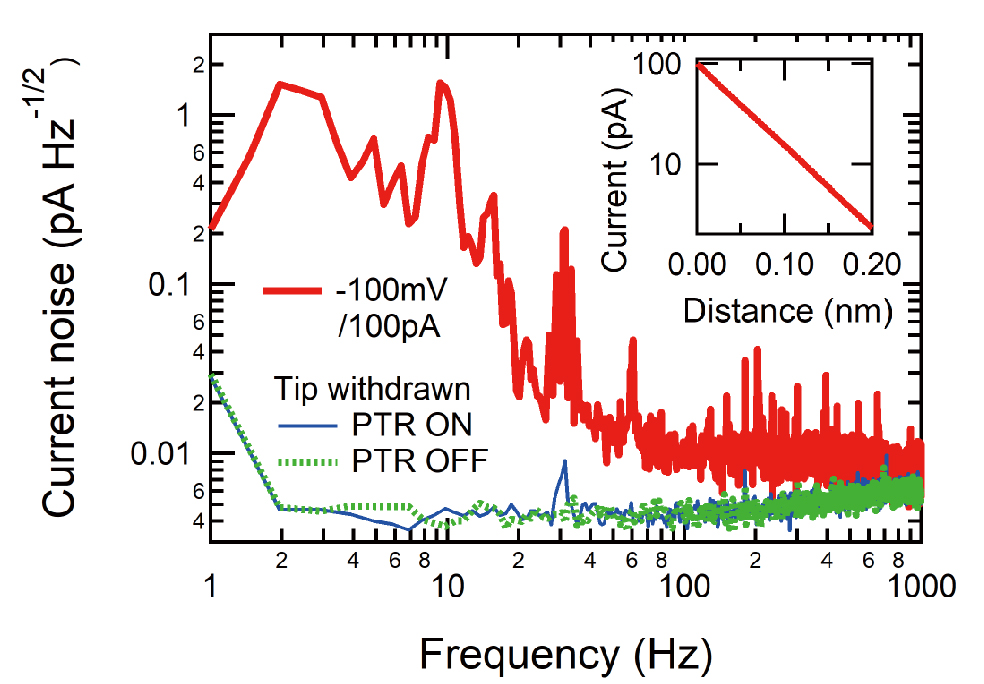}
\caption{\label{CurrentNoise} The tunneling-current noise spectrum measured on a Au(111) surface at $T = 7.0$~K with an open feedback loop (tunneling conditions: sample bias voltage, $V = -100$~mV; tunneling current, $I_{\rm t} = 100$~pA). The sample temperature was controlled to stabilize at 7.0~K. The inset shows the tunneling current as a function of the tip-sample distance. The light-solid (dotted) curve shows the tunneling-current noise spectrum when the tip is withdrawn from the sample surface with the PTR turned on (off).}
\end{figure}

To demonstrate the effects of vibration isolation on practical SPM measurements, we measure the tunneling-current noise spectra using a commercial current preamplifier ($10^9$~V/A, DLPCA-200, Femto Messtechnik GmbH, Germany), and a Nanonis SPM controller. 

Figure~\ref{CurrentNoise} compares the current noise spectra acquired with the tip withdrawn out of tunneling range from the sample surface with the PTR turned on and off (light-solid and dotted lines, respectively). The absence of any peak at 2~Hz during PTR operation indicates effective vibration isolation from the PTR. Although several peaks (e.g., at 30~Hz and 180~Hz) are observed, the peak heights are measurably low (below $10~\rm{fA}/\sqrt{\rm {Hz}}$). The noise floor remains low up to 1 kHz owing to the small capacitance afforded by the short length (50~cm) of the cable. This latter feature is advantageous for tunneling spectroscopy measurements using a lock-in amplifier.

We also show in Fig.~\ref{CurrentNoise} the tunneling-current noise spectrum acquired with a mechanically polished PtIr tip in tunneling range with an opened feedback loop over a Au(111)/mica surface prepared by repetitive Ar sputtering and annealing. The inset in Fig.~\ref{CurrentNoise} confirms the exponential decay of the tunneling current as a function of the tip-sample distance, ensuring an ideal tunneling condition. 

In the tunneling condition, a broad peak appears near 2~Hz, originating from the vibrations from both the PTR and the vibration isolation table. Several peak structures near 10, 15 and 30~Hz are attributed to the vibration isolation table as identified in Fig.~\ref{VibrationNoise2}. Although these noise peaks below 40~Hz are observed, the noise floor is measurably low for higher frequencies between 40~ Hz and 1~kHz.   

From the above discussions, we believe that a different relative configuration between the PTR and the SPM system may improve the noise level in the tunneling condition. In the current configuration shown in Fig.~\ref{SystemOverview}, the tip and sample are aligned in parallel with the cold head and valve unit in the x direction so that the tip-sample distance is most severely affected by the vibration noise from the PTR. Therefore, aligning the PTR in the y direction in Fig.~\ref{SystemOverview} while keeping the tip–sample direction unchanged could reduce vibrational coupling in the x direction. Another possibility would be to use an active vibration isolation table as reported in Ref.~\cite{Zhang}. Since the noise level on the active vibration isolation table in Ref.~\cite{Zhang} is substantially lower than in this study, its approach could further reduce noise peaks in the tunneling current.

\section{PERFORMANCE DEMONSTRATIONS}

%Fig.~\ref{SCgap}%
\begin{figure}
\includegraphics[width=7.5cm]{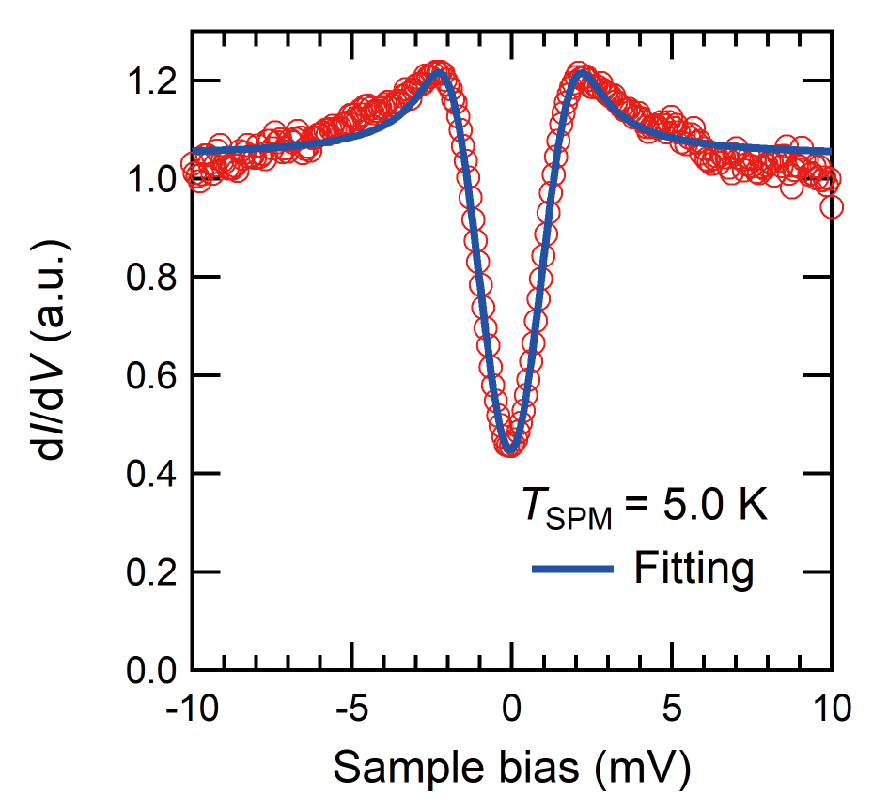}
\caption{\label{SCgap} A typical differential-conductance spectrum of Pb at the base temperature of 5.0~K, measured using the standard lock-in technique with an AC modulation amplitude $V_{\rm mod}$ of 212~$\mu V_{\rm rms}$ at 1 kHz (tunneling conditions: $I_{\rm t} = 500~{\rm pA}$, $V = +10~{\rm mV}$. The solid line shows the data fit using the Dynes function. Both thermal broadening and lock-in broadening are included.}
\end{figure}

First, we demonstrate the measurement of a Pb superconducting gap spectrum at $T = 5.0$~K and the evaluation of its effective electron temperature (Fig.~\ref{SCgap}).  Mechanically polished PtIr tips are used for all STM measurements in this study. We use a polycrystalline Pb plate (purity: 99.9\%) purchased from Nilaco Co., Ltd. To avoid RF noise propagating to the SPM head, we use low-pass filters for thermometry (500~kHz cutoff), a heater (2~MHz cutoff), XY and Z stage lines (100~kHz cutoff). In the scanning line, we add a capacitor of $\sim 10$~nF between the ground at the feedthrough of the chamber. All lines unused for the measurement are grounded. 
The superconducting gap spectrum is fitted with the Dynes function (see Appendix B about the detail), yielding the effective temperature of 5.7~K with the superconducting gap $\Delta = 1.29~{\rm meV}$ and a quasiparticle-lifetime broadening factor $\Gamma = 0.48~{\rm meV}$. 

%Fig.~\ref{STM}%
\begin{figure}
\includegraphics[width=8.5cm]{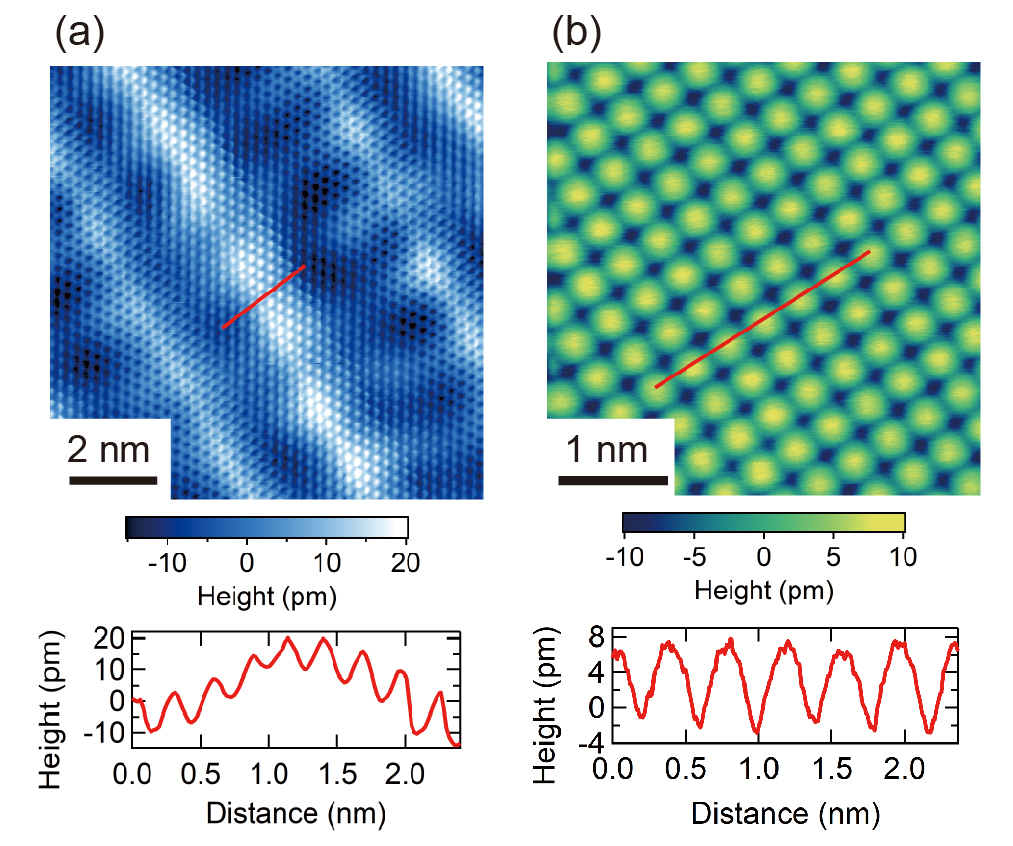}
\caption{\label{STM}  (a) Constant-current STM image of Au(111) on mica at $T = 5.6~{\rm K}$ (tunneling conditions: $I_{\rm t} = 1~{\rm nA}, V = +5~{\rm mV}$). (b) Non-contact AFM image of NaCl(100) at $T = 6.0~{\rm K}$ with a frequency shift $\Delta f = -13.2~{\rm Hz}$ at a resonance frequency $f_{0} = 29.29~{\rm kHz}$. $V = +1.7~{\rm V}$. The line profiles are shown under the corresponding images. The red line in each image indicates the location of the line profile. Both images are processed only by line subtraction.}
\end{figure}

Next, we demonstrate the constant-current STM imaging of a Au(111) surface, resolving its hexagonal atomic lattice along with the herringbone structure [Fig.~\ref{STM}(a)]. The atomic corrugation on a metallic surface such as Au(111) is considerably smaller compared to semiconducting and non-metallic surfaces so that a mechanically stable SPM is generally required to obtain atomic resolution. The line profile in Fig.~\ref{STM}(a) shows a well-defined atomic corrugation of $\sim 10$~pm without noticeable noise.

%Fig.~\ref{QPI}%
\begin{figure}
\includegraphics[width=8.5cm]{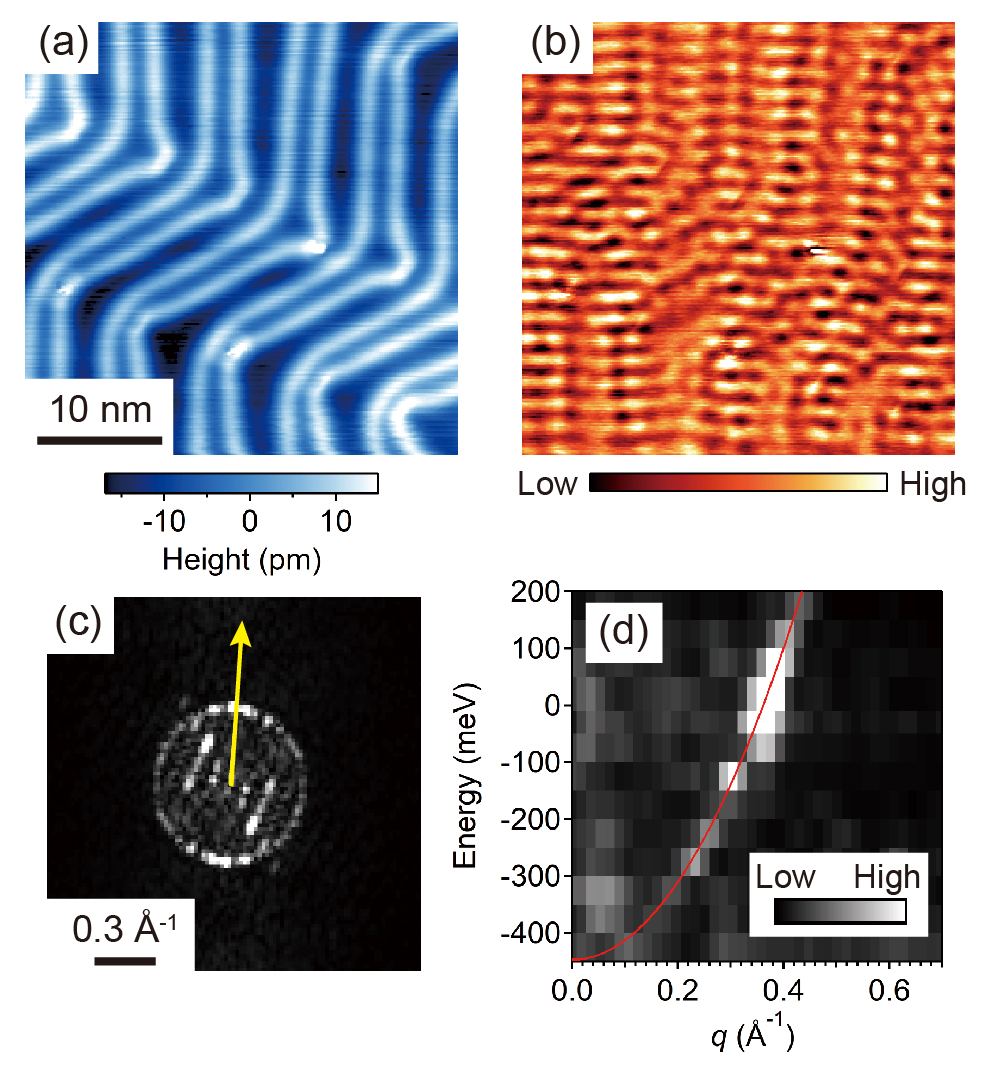}
\caption{\label{QPI} (a) Constant-current STM image of Au(111) on mica at $T = 7.0~{\rm K}$ (tunneling conditions: $I_{\rm t} = 1~{\rm nA}, V = -600~{\rm mV}$). (b) d$I$/d$V$ image at $V = -10~{\rm mV}$ acquired at the same location as in (a). $I_{\rm t} = 1~{\rm nA}$. $V_{\rm mod} = 7~{\rm mV_{rms}}$ at 1 kHz. (c) Fourier-transformed image of (b). (d) Energy dispersion of the QPI signal in the direction indicated in (c). The solid line serves as an eye guide. }
\end{figure}

Next, we demonstrate the non-contact AFM measurements on a NaCl(100) surface using the qPlus sensor\cite{Giessibl} [Fig.~\ref{STM}(b)]. The NaCl(100) single crystal (Goodfellow) is glued onto a silicon substrate and cleaved in air. We anneal the sample in UHV at about $100~^\circ$C for 15~min by a direct-current heating of the silicon substrate to avoid charging the surface. We use an electrochemically etched tungsten tip glued onto the free prong of the qPlus sensor. We mechanically excite the sensor using the inner piezo tube and the resonance frequency of the sensor is 29.29~kHz with a Q-factor of 60,000 at 6.0~K. The signal is amplified using a commercial charge amplifier (HQA-15M-10T, FEMTO Messtechnik) placed outside the observation chamber. For imaging, the tip-sample distance is controlled with a feedback loop to maintain a constant frequency shift $\Delta f = -13.2~{\rm Hz}$ while the tip is driven at a constant oscillation amplitude of 200~pm. The line profile in Fig.~\ref{STM}(b) shows a typical corrugation of approximately 10~pm, comparable to the spatial resolution of known low-temperature AFM systems cooled by liquid helium\cite{Bettac}.

Finally, we demonstrate the quasiparticle interference (QPI) imaging of a Au(111) surface. Figures~\ref{QPI}(a) and (b) show the topographic and the differential-conductance (d$I$/d$V$) images acquired simultaneously on Au(111), respectively. The interference pattern caused by electrons scattering at surface defects are clearly observed in Fig.~\ref{QPI}(b). The d$I$/d$V$ image in Fig.~\ref{QPI}(b) is acquired using a lock-in amplifier at $\sim13$~min imaging time. We use Fourier analysis to identify QPI signals appearing at scattering wave $q$ vectors. The Fourier-transformed d$I$/d$V$ image in Fig.~\ref{QPI}(c) exhibits a ring-like feature around $q = 0$, reflecting the isotropic backscattering process within the ring-like constant energy contour of the parabolic surface state centered at the $\Gamma$ point\cite{Petersen}. By varying the bias voltage, the QPI signal shows a parabolic energy dispersion [Fig.~\ref{QPI}(d)], as expected for the two-dimensional surface state. The dispersion is in good agreement with previously reported data\cite{Hoesch, Tesch}. We note that thirteen d$I$/d$V$ images are used to compile the data shown in Fig.~\ref{QPI}(d) which requires a minimum of 3 hours uninterrupted acquisition time. Without the need for cryogen refills, the current SPM system can, in principle, provide uninterrupted measurement time indefinitely.

\section{CONCLUSION}

This work reports the design of a cryogen-free, PTR-based UHV SPM system capable of  operating at the base temperature of $\sim 5$~K. We achieve this by combining a home-made interface design between the PTR cold head and the SPM head with an automatic gas-handling system: while the interface design enhances both vibration isolation and cooling capability, the gas-handling system facilitates both the cool down to and the maintenance of the base temperature. Our report provides both the detailed information on the vibration isolation structures, and the corresponding vibration noise measurements. Our SPM data show comparable quality with those acquired by conventional cryogen-cooled instruments. Our work demonstrates the possibility for achieving delicate low-temperature measurements without cryogenic liquids.

\begin{acknowledgments}
We are grateful to Yo Miyazaki, Tetsuya Maeda, Nozomi Nishiyama, Yutaka Miyatake (UNISOKU) and Tokkyokiki Corporation for their technical assistance. We would also like to thank Tadashi Machida (RIKEN) for helpful discussion. This study was partly supported by the Small and Medium Enterprise Agency (Grant No.~2527213203). 
\end{acknowledgments}

\section*{AUTHOR DECLARATIONS}
\subsection*{Conflict of Interest}
The authors have no conflicts to disclose.

\section*{DATA AVAILABILITY}
The data that support the findings of this study are available from the corresponding author upon reasonable request.

\appendix
\section{DOUBLE PIEZO TUBE STRUCTURE}
\begin{figure}
\includegraphics[width=7cm]{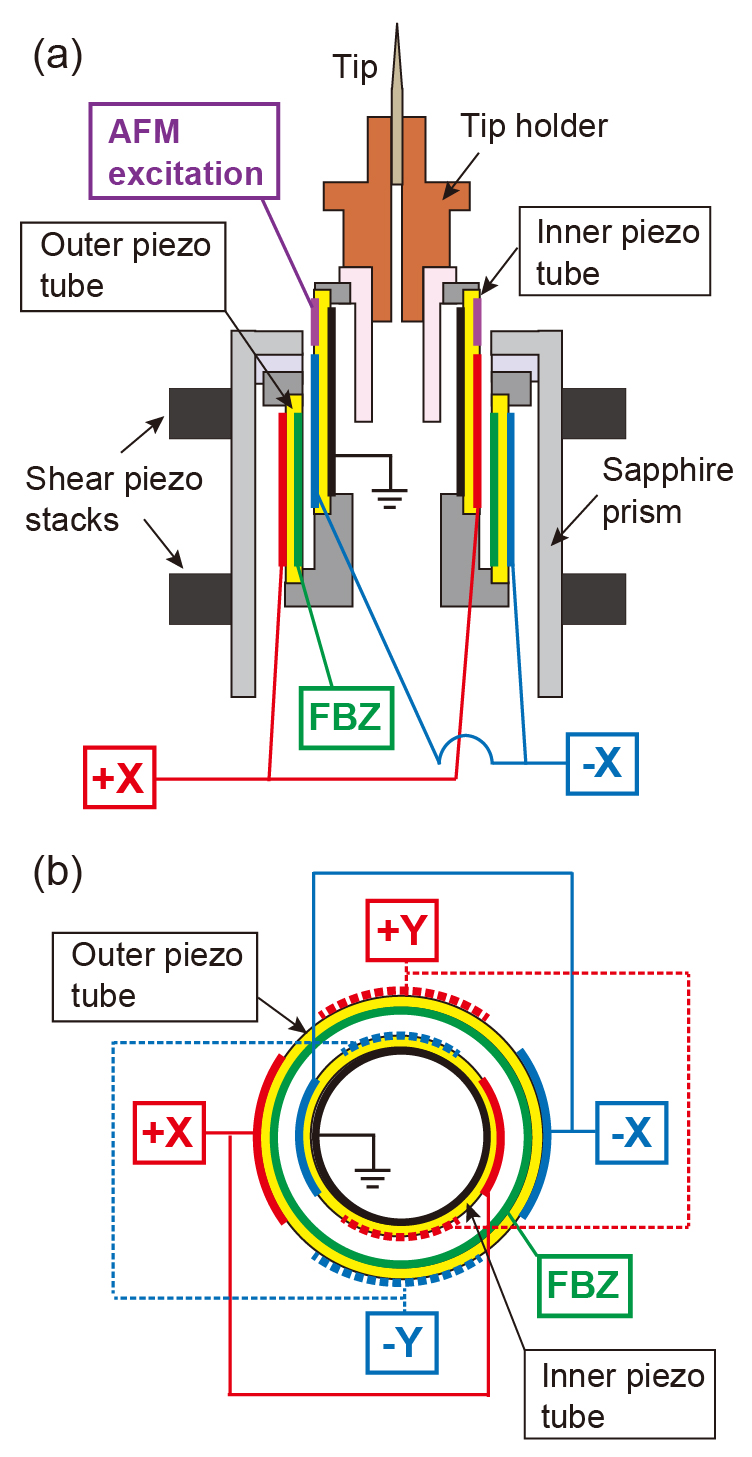}
\caption{\label{Piezo} Schematic of the double piezo tube structure. (a) Side-view. The STM tip holder is shown. (b) Bottom-view.}
\end{figure}

The configuration of electrodes in the double piezo tube structure is shown in Fig.~\ref{Piezo}. The top of the outer piezo tube is fixed to the sapphire prism, and the bottom end of the inner piezo tube is fixed to the bottom of the outer piezo tube [Fig.~\ref{Piezo}(a)].  
Therefore, in order to scan the tip laterally in a wide range (for instance, $1.5~\rm{\mu m} \times 1.5~\rm{\mu m}$ at low temperatures), the inner and outer piezo tubes need to be deflected in an opposite direction. To realize such a deflection, a high voltage is simultaneously applied to an outer quadrant electrode of the outer piezo tube and also to an outer quadrant electrode located on the opposite side of the inner piezo tube [Figs.~\ref{Piezo}(a,b)].    
The upper part of the outer electrode of the inner piezo tube is used for mechanically exciting the qPlus sensor. 
The inner electrode of the outer piezo tube is assigned for feedback Z (FBZ) motion whereas the inner electrode of the inner piezo tube is grounded.

%Fig.~\ref{Piezo}%

\section{FUNCTION FOR FITTING THE TUNNELING SPECTRUM OF LEAD}
To estimate the effective temperature $T_{\rm eff}$, the tunneling spectrum of Pb shown in Fig.~\ref{SCgap} is fitted with a theoretical formula $g(V)$, 
\begin{equation}
    g(V)=-\int^\infty_{-\infty}\left\lbrace \int^\infty_{-\infty} \rho(E)f'(\epsilon + E)dE\right\rbrace b(eV-\epsilon)d\epsilon
\end{equation}
where $e$ and $V$ is the elementary charge and the bias voltage, respectively. The Dynes function\cite{Dynes} representing the density of states of a superconductor $\rho(E)$ is convolved with the energy derivative of the Fermi function $f'(\epsilon)$ and the lock-in broadening function $b(V)$, 
\begin{equation}
    \rho(E)={\rm Re}\left( \frac{E-i\Gamma}{\sqrt{(E-i\Gamma)^2-\Delta^2}} \right)
\end{equation}

\begin{equation}
    f'(\epsilon)=\frac{df(\epsilon)}{d\epsilon}=-\frac{\beta\exp(\beta\epsilon)}{\lbrace\exp(\beta\epsilon)+1\rbrace^2}
\end{equation}

\begin{align}
    b(V) &=
    \begin{cases}
        \frac{\sqrt{2}}{\pi V_{\rm mod}}\sqrt{1-\left( \frac{V}{\sqrt{2}V_{\rm mod}} \right)^2} & (|V| \leqq \sqrt{2}V_{\rm mod})\\
        0 & (|V| > \sqrt{2}V_{\rm mod})
    \end{cases}
\end{align}
where $\beta=(k_{\rm B}T_{\rm eff})^{-1}$ and $\Delta$, $\Gamma$, and $V_{\rm mod}$ is the superconducting gap amplitude, the quasiparticle-lifetime broadening factor, and the root-mean-square amplitude of lock-in excitation, respectively.

%\nocite{*}
%\bibliography{References}

\end{document}